\documentclass[11pt]{article}
\usepackage{graphicx}
\usepackage{amssymb,amsmath}%
\usepackage{hyperref}
\begin{document}

\title{Spatial distribution of spin-wave modes in cylindrical nanowires of finite aspect ratio}

\author{V. O. Dolocan$^1$\\
\small $^1$Aix-Marseille Univ \& IM2NP CNRS UMR 6242\\[-0.8ex]
\small Avenue Escadrille Normandie Niemen, 13397 Marseille, France\\
\small email: \texttt{voicu.dolocan@im2np.fr}
}
\date{} 


\maketitle

\begin{abstract}
The spin wave modes of cylindrical nanowires of moderate diameter-to-length ratio are investigated in this article. Based on three dimensional simulations and analytical calculations we determine the spatial structure of the modes. We show that standing spin waves and localized edge modes form the discrete spectrum of the nanowires. Using a simple analytical model we infer an extended dispersion relation for spin waves in cylinders. Considering the variation of the demagnetizing (internal) field we show that the localized dipole-exchange modes at the edges are always present.

\end{abstract}


\section{Introduction}

Magnetic nanostructures are intensively studied nowadays for fundamental and practical purposes\cite{Hillebrands2001.vol.1}. Understanding the magnetism in low-dimensions presents a fundamental interest, while achieving smaller data storage media and magnetic memory is of utmost technological importance. Large periodic arrays of magnetic nanowires have been easily produced with inexpensive techniques like electrodeposition\cite{Pirota200418}. For such structures, the manipulation of magnetization in very short times is very important for high-speed applications and requires a complete understanding of the nature of magnetic excitations, magnons or spin-waves, and their dependence on geometry.   

The spin-wave spectrum for ellipsoidal samples is known for some time\cite{PhysRev.105.390,PhysRev.110.1295}. In uniform magnetized ellipsoidal samples the demagnetizing field is uniform. As many of the magnetic nanostructures studied nowadays have in general a non ellipsoidal form, the demagnetizing field inside them is nonuniform even if they are uniformly magnetized\cite{Joseph}. Thus the shape can drastically affect the dynamic properties and the spectrum of the spin-waves will be modified (see Ref.\cite{Demokritov2008} and references therein). As the dimension of the wires decreases, the exchange interaction becomes important. We consider here that both dipolar and exchange interactions contribute to the spin-wave spectrum. Obtaining a general theory is challenging, but approximate analytical solutions and numerical results can be derived for certain geometries. 

Reducing the dimensions of magnetic nanostructures is the natural trend to obtain smaller devices. Until now, only very long nanowires were studied with diameter-to-length ratio (d/L) $\ll$ 1. In this paper, we investigate nanowires of moderate aspect ratio magnetized along the axial direction. We develop the analytical theory and present numerical results of the spatial distribution of spin waves modes in this type of nanostructures. We will outline the important features of taking into account the in-homogeneity of the demagnetizing field for the confined geometry under study. The previous studies of cylindrical geometry neglected the nonuniformity of the demagnetizing fields considering the cylinders semi-infinite\cite{Joseph2,PhysRevB.63.134439}.

\section{Analytical model}

We start with a ferromagnetic cylindrical sample, magnetized to saturation along the axial direction (\textit{z} direction) which ensures that the longitudinal component of \textbf{M} is much larger than the transverses ones, \textbf{M} = ($m_{x}, m_{y}$, M$_{s}$). A static magnetic field \textbf{H} is applied along z and an rf field \textbf{h}(\textbf{r},t)=\textit{h}(\textbf{r})e$^{-i\omega t}$ is applied in perpendicular direction, $\textbf{H}_{ext} = (h_{x}, h_{y}, H)$. We assume that the rf field and the saturation magnetization M$_{s}$ are uniform in the sample, even if the state of uniform magnetization is not actually realized in this geometry. Two main approaches are used for solving the spectrum of spin-waves in confined structures. In both, one solves simultaneously the linearized Landau-Lifschitz equation of motion for the magnetization together with the Maxwell equations satisfying the electromagnetic boundary conditions and the exchange boundary conditions\cite{Kalinikosbook}. The two methods are equivalent. We choose here the method of magnetic potential where the relation \textbf{m}(\textbf{h}) is found first from the equation of motion and then we search for the solutions of Maxwell equations which satisfy the boundary conditions.

The Landau-Lifschitz (LL) equation of motion for the magnetization neglecting damping is:

\begin{equation}
\label{eq1}
\frac{d\textbf{M}}{dt}(\textbf{r},t) = -\gamma\textbf{M}\times\textbf{H}_{eff}
\end{equation}

\noindent where $\gamma$ is the gyromagnetic ratio. The effective field inside the sample represents the sum of the external applied field $\textbf{H}_{ext}$, the exchange field $\textbf{H}_{exch}=D\nabla^{2}\textbf{M}$ and the demagnetizing field $\textbf{H}_{demag}= - \widehat{\textbf{N}}\textbf{M}$ (excluding crystal anisotropy):

\begin{equation}
\label{eq2}
\textbf{H}_{eff}=\textbf{H}_{ext}+\textbf{H}_{exch}+\textbf{H}_{demag}
\end{equation}

\noindent where D is the exchange stiffness and $\widehat{\textbf{N}}$ the demagnetizing tensor field. For non-ellipsoidal bodies the demagnetizing tensor field (in the first order) is function of position and is defined in the Fourier space as in Ref.\cite{Tandon20049}:

\begin{equation}
\label{eq3}
N_{ij}(\textbf{k}) = \frac{D(\textbf{k})}{k^2}k_ik_j
\end{equation}

\noindent with D(\textbf{k}) the shape function. The LL equation (Eq.(\ref{eq1})) describe an uniform precession of the magnetization around the effective field $\textbf{H}_{eff}$. To find the normal modes of the magnetization we search for non zero solutions of the LL equation. Considering the dynamical magnetization uniform in the sample we find:

\begin{align}
\label{eq4}
i\omega m_x &= \gamma m_y[H_i-D\nabla^2]-\gamma M_sh_y \nonumber \\ 
-i\omega m_y &= \gamma m_x[H_i-D\nabla^2]-\gamma M_sh_x 
\end{align}

\noindent with $N_{zz}$ and $N_{rr}$ the longitudinal (dc) and transversal (ac) demagnetizing factors and H$_{i}$=H--4$\pi$(N$_{zz}$--N$_{rr}$)M$_{s}$. Here, we'll consider as first order approximation that the demagnetizing tensor field can be diagonalized as $N_{zz}$+2$N_{rr}$=4$\pi$. 

The demagnetizing field is determined by Maxwell equations in the magnetostatic limit:

\begin{align}
\nabla\times\textbf{H}_{int} & = 0 \label{eq5}\\ 
\nabla \textbf{B}=\nabla(\textbf{H}_{int} + 4\pi\textbf{M}) & = 0 \label{eq6}
\end{align}

\noindent where the internal field is $\textbf{H}_{int}=\textbf{h}-N_{rr}\textbf{m}+H-N_{zz}M_s$. From Eq.(\ref{eq5}), the magnetic field may be written as $\textbf{h} = -\nabla\Psi_m$, where $\Psi_m$ is the scalar magnetic potential. Replacing the scalar potential in Eq.(\ref{eq6}) we obtain:

\begin{equation}
\label{eq7}
-\nabla^2\Psi_m+(4\pi-N_{rr})(\frac{\partial m_x}{\partial x}+\frac{\partial m_y}{\partial y})-M_s\frac{\partial N_{zz}}{\partial z}=0
\end{equation}

Eq.(\ref{eq7}) has the form of an anisotropic Laplace equation or Walker type equation for cylinders with the last term coming from the variation of the demagnetization (internal) field near edges. In the center of the cylinder, where the demagnetization field is almost constant (here the longitudinal demagnetization field is zero) the Walker equation reduces to the classical equation for cylinders\cite{PhysRevB.63.134439}. We considered here only the variation of the longitudinal demagnetizing factor $N_{zz}$ along z (see Appendix for more details). The nonuniformity of $N_{zz}$ strongly affects the mode frequencies as each mode feels a different demagnetizing factor. The full analysis should also take into account the variation of N$_{rr}$ in the radial direction r, which we neglect here as Nrr is almost constant with r due to the fact that our cylinders have a reduced aspect ratio (diameter-to-length inferior of 0.1) and such an analysis will introduce a high degree of analytical complexity. 


\begin{figure}[t]
\center
  \includegraphics[width=7cm]{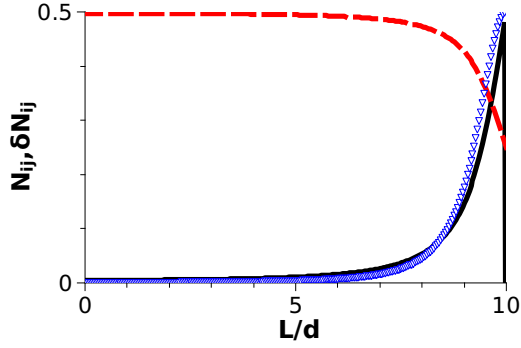}
  \caption{\label{Fig.1} Variation of the demagnetizing tensor components (here normalized to $N_{zz}$+2$N_{rr}$=1) along the axial direction for a cylinder with length-diameter (L/d) ratio of 10: N$_{zz}$ solid line, N$_{rr}$ dashed line and $\partial$N$_{zz}$/$\partial$z triangles. Only half of the cylinder is shown with largest variation near the edge.}
\end{figure}


We state that we made here the simplifying assumptions: the static magnetization is considered as constant and independent of the external field and that the dc and ac demagnetizing fields are the same and can be calculated with the expressions given in Eq.(\ref{eq3}). Even if the non-diagonal elements of the demagnetizing tensor field are not zero, they can be overlooked as they are negligible (the N$_{rz}$ component is inferior of 10$^{-4}$). The variation of the demagnetizing factors\cite{Tandon20049} along the axial direction is shown in the Fig.\ref{Fig.1}. 

Choosing the scalar potential of the form $\Psi_m(r,\varphi,z)$=J$_{n}$(qr)exp(in$\phi$+ik$_{z}$z) where J$_{n}(qr)$ is the Bessel function of order \textit{n}, we obtain the transcendental equation (see Appendix):

\begin{multline}
\label{eq8}
\omega^{2}(k_{z}^{2}+q^{2}-M_{s}\frac{\partial N_{zz}}{\partial z})=\gamma^{2}D^{2}(k_{z}^{2}+q^{2})^{3}+\gamma^{2}D(k_{z}^{2}+q^{2})^{2}\\
\times\left[ 2H_{i}+(4\pi-N_{rr})M_{s}\right]+\gamma^{2}(k_{z}^{2}+q^{2})H_{i}\left[ H_{i}+(4\pi-N_{rr})M_{s}\right]  \\ 
-k_{z}^{2}\gamma^{2}M_{s}(4\pi-N_{rr})H_{i}-\gamma^{2}M_{s}D(4\pi-N_{rr})k_{z}^{2}(k_{z}^{2}+q^{2})\\
-\frac{\partial N_{zz}}{\partial z}\gamma^{2}M_{s}H_{i}^{2}  +\frac{\partial^{3}N_{zz}}{\partial z^{3}}2\gamma^{2}M_{s}DH_{i}-\gamma^{2}D^{2}M_{s}\frac{\partial^{5}N_{zz}}{\partial z^{5}}
\end{multline}

In the limit of very long wavelength, when the wave vector has all components zero we derive the following expression:

\begin{multline}
\label{eq9}
-\frac{\partial N_{zz}}{\partial z}\omega_0^{2}=-\frac{\partial N_{zz}}{\partial z}\gamma^{2}H_{i}^{2}  +\frac{\partial^{3}N_{zz}}{\partial z^{3}}2\gamma^{2}DH_{i}-\\
\gamma^{2}D^{2}\frac{\partial^{5}N_{zz}}{\partial z^{5}}
\end{multline}

\noindent which takes the form of the Kittel uniform mode\cite{PhysRev.71.270.2} if we neglect the higher exchange terms (derivative of order three and five) and keep only the first derivative term: $\omega_0=\gamma H_{i}$. However, no real uniform mode exists in cylinders of finite aspect ratio as the internal field varies near edges. In the case of a nonuniform magnetic field we usually assume that a spin wave can propagate with continuously changing wave vector $k_z$(z)\cite{Schlomann}. At certain points (surface), the wave vector becomes zero and the condition of quasi-static variation of the internal field is no longer satisfied thus the solution of the equation of motion has to be found with variable parameters. This surface is called turning surface by analogy with quantum mechanics. The Eq.(\ref{eq9}) at the turning surface has no physical meaning. A localized spin wave is excited with changing wave vector propagating from the turning surface in the direction of the decreasing internal field. This effect called spin-wave well was demonstrated in stripes recently\cite{PhysRevLett.88.047204}.

For samples of finite size, the dispersion relation is shape dependent. The complete spin-wave spectrum and the modes shape depend strongly on the boundary conditions. We expect some degree of pinning at the edges, and this degree of pinning is determined by a competition between dipolar magnetostatic energy and exchange energy. Apart from the Maxwell boundary conditions, continuity of the magnetization and the magnetic potential, usually a Rado-Weertmann type of boundary condition\cite{JPhysChemSolids.11.315,PhysRevB.72.014463} is considered at the cylinder surface:

\begin{equation}
\label{eq10}
\frac{\partial m}{\partial z}\pm p\times m=0
\end{equation}

\noindent at z=$\pm$L/2 with L the length of the cylinder and p the so-called pinning parameter which is determined by the effective surface anisotropy K$_{s}$ and the exchange stiffness constant D: p$\simeq$K$_{s}$/D. This condition implies stationary waves in the z direction with the particular solution of the type $m_{z}$=A$\sin k_{z}z$+B$\cos k_{z}z$.

The frequency of an eigenmode is constant throughout the magnet and the spin wave vector varies to accommodate the changes in the internal field profile. The dynamic magnetization has a plane-wave character. In almost all the volume of the cylinder, between the turning surfaces at both ends (excluding the edge domains), the variation of the demagnetization field is small and usually the averaged value of the components of demagnetizing field is used. The derivatives of the  N$_{zz}$ in the Eq.(\ref{eq8}) are negligible. For moderate aspect ratio cylinders we can use a simple and intuitive model where the dynamical magnetization takes the form:

\begin{equation}
\label{eq11}
m_{ln}(r,z) \sim J_{n}(qr)\mbox{cos}(k_{z}^{l}z),
\end{equation}

\noindent where k$_{z}^{l}$ is the longitudinal wave number for the \textit{l}th longitudinal mode (backward geometry) and \textit{n} indexes the radial modes. This relation is valid between the turning surfaces, where the internal field is almost constant. In this region, stationary waves on the axial and radial directions are formed and we can have mixed modes. The distance between the turning surfaces, or effective length $\Delta$z, is considered to be approximately 0.8L (with L the length of the cylinder). The dispersion curves are characterized by two quantization numbers \textit{n} and \textit{l}. The quantization parameter \textit{l} should be taken with care, it provides here a qualitative description of longitudinal modes which is to be compared with the micromagnetic simulations. We can use the mean value of the wave vector of each mode which can be evaluated as: k$_{z}^{l}=l\pi$/$\Delta$z, where $\Delta$z is the effective length where the mode is localized in the sample. The spectrum of the SW consists of a series of dispersion curves characterized by two quantization numbers \textit{n} and \textit{l} and depend on the ratio of the cylinder dimensions. Changing the aspect ratio of the cylinders induces a change in the values of wavenumbers (and also a change in the demagnetizing field). For example, if the aspect ratio is diminished, q becomes much larger than k$_{z}$. The principal effect is a reduced influence of  k$_{z}$ on frequency or applied field and thus the longitudinal modes form a continuous band. When d/L=0.1, the difference between the first and the fifth longitudinal mode is of 0.8kOe at fixed frequency (20GHz). The same difference is only of 0.03kOe when d/L=0.02, while the difference in the radial modes is much larger.


\subsection{Spatial distribution of edge modes}

\begin{figure}[t]
\center
  \includegraphics[width=7cm]{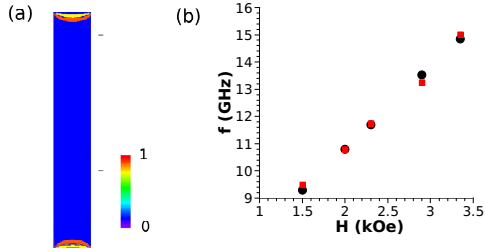}\\
  \caption{\label{Fig.2} (a) Spatial distribution of an edge mode calculated from the analytical expression for a cylinder with diameter-to-length ratio of 0.1 (L=300nm) in a false color code. Red (positive) represents a permitted value for the wavenumber. (b) Edge mode frequencies determined from the analytical expression (circles) and from micromagnetic simulation (rectangles) at different DC magnetic fields. }
\end{figure}

The terms which contain the derivatives of the demagnetizing field that appear in the Eq.(\ref{eq8}) have an increased influence near the edges of the cylinder where their variation is most important. These terms will provide a correction to the frequencies of the edge modes. To quantify this correction we numerically calculated the frequencies of the edge modes for a Nickel cylinder of aspect ratio d/L=0.1 (L=300nm) with and without the additional terms at fixed magnetic field. For example, without correction we obtain a frequency of 13.79GHz for the edge mode at 1.5kOe, a 48\% difference from the 9.3GHz value obtained with the correction terms. The frequencies obtained for the edge modes by the analytical formula with the correction terms are represented in the Fig.\ref{Fig.2}(b) along with the values obtained from micromagnetic simulations discussed in the next section. These values are similar which imply that the correction terms are necessary to calculate exactly the frequencies of edge modes in finite size nanowires.

In Fig.\ref{Fig.2}(a), we show the simulated 2D spatial distribution of the edge mode near surface for a DC magnetic field of 2.9kOe. The spatial distribution is obtained from the permitted values of the wavenumber near surface (positive). Our method is the following: we first numerically determine the frequencies of the edge modes near boundaries assuming that it does not propagate into the center of the cylinder. Next, we fix the frequency at the value obtained in the first step (at a given DC magnetic field) and we calculate the spatial distribution of the allowed values for the wavenumber in the edge region. From the plot we observe that the edge mode is not restricted to the actual surface but goes underneath the surface in parabolic fashion.


\section{Micromagnetic simulation}

In order to understand the mode structure and to compare the analytical results of the previous section, a micromagnetic simulation was conducted with the nmag package\cite{Fischbacher} for a cylinder with an aspect ration d/L=0.1 (L=300nm). The cylinder was discretized with a cell size of 5nm. First, the magnetization was relaxed to equilibrium using a large damping parameter $\alpha$=0.5. The equilibrium configuration was then excited adiabatically with a small 20GHz rf magnetic field in the perpendicular direction, while a constant magnetic field was applied along the cylinder axis. The values used for the parameters are the same as used in the analytical simulation and are those of Ni: D=2$\times$10$^{-9}$Gcm$^{2}$, $\gamma$=188.5GHz/T (g factor of 2.15) and M$_{s}$=480 emu/cm$^{3}$\cite{PhysRevB.63.134439,PhysRevB.64.144421} and a damping parameter of 0.015. We subsequently computed the spatial distribution of the oscillation of the dynamic magnetization doing Fourier transforms on a number of cells along the z and x axes of the cylinder. For comparison with previous results, we kept fixed the frequency of the rf magnetic field and we varied the amplitude of the DC magnetic field. 

\begin{figure}[t]
\center
  \includegraphics[width=7cm]{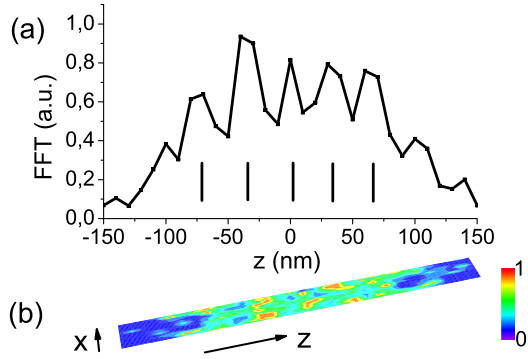}\\
  \caption{\label{Fig.3} (a) Average magnetization profile along z axis of a cylinder with diameter-to-length ratio of 0.1 (L=300nm). The average was calculated from 30 magnetization profiles taken each nm in the x direction in the plane y=0; (b) 2D color projection in the xz plane of the 30 magnetization profiles. The maximal amplitude of the dynamical magnetization (red) precess at 20GHz with a DC magnetic field applied (2.9kOe) corresponding to the 5th longitudinal mode.}
\end{figure}

To determine the spatial distribution of the longitudinal modes in an cylindrical nanowire we calculated the amplitude of oscillation of the magnetization at different magnetic fields. The values of the static magnetic field were chosen to correspond to longitudinal modes as calculated with Eq.(\ref{eq8}).  In Fig.\ref{Fig.3}(a), we show the average of the magnetization profiles along the z-axis for one DC magnetic field which should correspond to the 5th longitudinal spin wave mode determined from the analytical model. These profiles are calculated doing Fourier transforms on the magnetization for each nm in the z direction and in the x direction and extracting the values at 20GHz. Basically, we compute the amplitude of the dynamical magnetization variation in a xz-plane passing through the center of the cylinder (Fig.\ref{Fig.3}(b)) and then we average between the curves in the x direction. This corresponds with the average of a contour plot in a xz-plane. As observed, the calculated profile corresponds well with the one expected at this applied external field from the analytical model. The spins which form the longitudinal mode precess are mainly in the effective length $\Delta$z. Another mode appear at the edges of the cylinder where the spins precess at a lower frequency.


\begin{figure}[t]
  \centering
  \includegraphics[width=7cm]{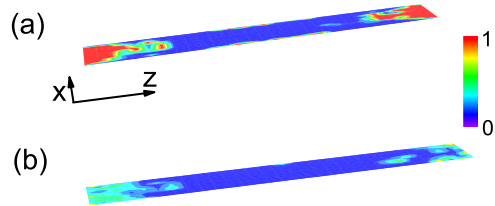}\\
  \caption{\label{Fig.4} 2D color projection in the xz plane of two edge modes of a cylinder with diameter-to-length ratio of 0.1 (L=300nm) at two different DC magnetic fields: (a) 1.5kOe and (b) 2.9kOe.}
\end{figure}

In Fig.\ref{Fig.4}, the 2D spatial distribution of the magnetization is shown corresponding to the edge modes obtained at two different DC magnetic fields. The xz plane shown passes through the center of the cylinder, rotating it 360$^{\circ}$ around its axis will provide the 3D profile of the mode. The panel (b) corresponds to an edge mode with spins precessing at a lower frequency of 13.25GHz (DC field 2.9kOe) than the longitudinal mode shown in the Fig.\ref{Fig.3}(b) where the spins precess at 20GHz. The spatial distribution of edge modes comes into agreement with that calculated in the previous section (Fig.\ref{Fig.2}(a)). The frequencies of the edge modes obtained from the microwave simulation are shown in the Fig.\ref{Fig.2}(b). The values are similar with those calculated with the Eq.(\ref{eq8}).


\begin{figure}
\center
  \includegraphics[width=7cm]{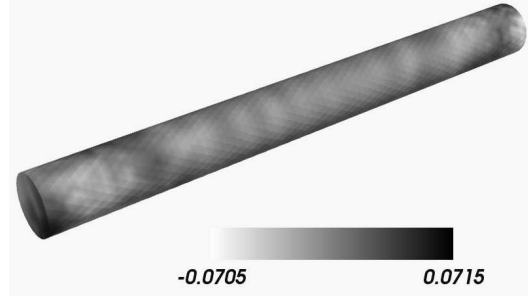}
  \caption{\label{Fig.5} 3D profile of the evolution of the y component of magnetization is shown at a given instant in time (30ps). The applied DC magnetic field is 2.9kOe and the frequency of the rf field is 20GHz.}
\end{figure}

In Fig.\ref{Fig.5}, a snapshot of the 3D profile of the y component of magnetization is shown after 30ps. We clearly observe the variation of the dynamical magnetization along the z direction (long axis). The mode is 3D longitudinal mode as expected even though some hybridization effects may exists, the standing waves picture being simplistic\cite{PhysRevB.70.054409}. The edge mode is clearly seen, it corresponds to the spins rotating at the surface (in plane) and precessing at lower frequency than the bulk modes. The edge mode is localized and it is always present due to the magnetic field inhomogeneity at the boundaries. The longitudinal modes which precess at 20GHz decay in time, becoming zero after 1.5ns.


\section{Discussion and conclusion}

The spatial distribution of longitudinal and edge modes in cylinders of moderate aspect ratio was calculated in this paper. It provides new insight on the spatial distribution of spin waves in cylindrical nanowires. We used a 2D analytical model and obtained the spin wave modes frequencies and profiles for the edge modes. The validity of the 2D model was tested through comparison with 3D micromagnetic results. The values of the resonances agree quantitatively although the profiles of the edge modes from 3D micromagnetics are not exactly parabolic and seem to penetrate on a longer distance beneath the surface (30-50nm). The difference between the two results are probably due to the assumptions and approximations that where made: in the analytical model the magnetization is considered uniform in the whole cylinder and a correction in the z direction is obtained but the variation of the radial demagnetizing field is not considered (which can be an useful extension of the theory) while in the 3D micromagnetics we break the cylinder into small tetrahedral cells (discretization length of 4.8nm) and consider that the magnetization is continuous and vary linearly and then the demagnetizing field is computed.

The nanowires used in the simulation have finite diameter-to-length ratio (d/L=0.1) and thus the longitudinal modes are well separated and their spatial distribution was resolved. We tried to resolve the spatial distribution of longitudinal modes for a cylinder with d/L=0.02 without success, as the modes are hybridized (closer in frequency) and form a quasi-continuum band. Furthermore, as the demagnetizing field is inhomogeneous it confines the spin waves at boundaries of the cylinder: the edge modes are always present in finite aspect ratio cylinders. These modes can be observed in experiments. Until now, only modes of an ensemble of nanowire have been investigated\cite{PhysRevB.64.144421}. To measure the spectrum of an individual nanowires a particularly well suited technique is the (Ferro)Magnetic Resonance Force Microscopy\cite{Zhang, Klein}. This local probe technique can validate our results.


\section*{Acknowledgments}
I would like to thank O. Klein, V. Dolocan, M. Franchin, H. Fangohr for their help with the simulations and L. Raymond and J.M. Debierre for use of the Lafite server.


\appendix
\section*{Appendix}
\setcounter{section}{1}

Noting $m_+=m_x+im_y$ et $m_-=m_x-im_y$ (similarly to the transverse right and left handed magnetization density operators) we can rewrite Eq.(\ref{eq7}) as:

\begin{multline}
\label{A1}
\nabla^2\Psi_m-\frac{4\pi-N_{rr}}{2} \left[ (\frac{\partial}{\partial x}-i\frac{\partial}{\partial y})m_+ +(\frac{\partial}{\partial x}+i\frac{\partial}{\partial y})m_- \right] \\
+M_s\frac{\partial N_{zz}}{\partial z}=0
\end{multline}

\noindent Applying the operators $(\frac{\partial}{\partial x}-i\frac{\partial}{\partial y})m_+$ and $(\frac{\partial}{\partial x}+i\frac{\partial}{\partial y})m_-$ to Eq.(\ref{eq5}), it can be easly shown that:

\begin{multline}
\label{A2} 
[\omega^2-\gamma^2(H_i-D\nabla^2)^2-(4\pi-N_{rr})\gamma^2M_s(H_i-D\nabla^2)]\nabla^2\Psi_m \\
+[\omega^2-\gamma^2(H_i-D\nabla^2)^2]M_s\frac{\partial N_{zz}}{\partial z}+(4\pi-N_{rr})\gamma^2M_s \\
\times(H_i-D\nabla^2)\nabla^2_{\perp}\Psi_m=0
\end{multline}

\noindent with $\nabla^2_{\perp}=\nabla^2-\frac{\partial^2}{\partial z^2}$.

Replacing the scalar potential into Eq.(\ref{A2}) we obtain Eq.(\ref{eq8}).



\begin{thebibliography}{18}

\bibitem{Hillebrands2001.vol.1} Hillebrands B and Ounadjela K 2006 {\it Spin Dynamics in Confined Magnetic Structures} vol 3 (Berlin, Heidelberg, New York: Elsevier)
\bibitem{Pirota200418} Pirota K R, Navas D, Hern\'{a}ndez-V\'{e}lez M, Nielsch K and V\'{a}zquez M 2004 {\it Journal of Alloys and Compounds} {\bf 369} 18
\bibitem{PhysRev.105.390} Walker L R 1957 {\it Phys. Rev.} {\bf 105} 390.
\bibitem{PhysRev.110.1295} Kittel C 1958 {\it Phys. Rev.} {\bf 110} 1295.
\bibitem{Joseph} Joseph R I and Schlomann E 1965 {\it J. Appl. Phys} {\bf 36} 1579
\bibitem{Demokritov2008} Demokritov S O 2008 {\it Spin Wave Confinement}(Singapore: Pan Stanford Publishing)
\bibitem{Joseph2} Joseph R I and Schlomann E 1961 {\it J. Appl. Phys} {\bf 32} 1001
\bibitem{PhysRevB.63.134439} Arias R and Mills D L 2001 {\it Phys. Rev. B} {\bf 63} 134439
\bibitem{Kalinikosbook} Kalinikos B A 1994 {\it Linear and Nonlinear Spin Waves in Magnetic Films and Superlattices} ed M G Cottam (New York: World Scientific) p~89
\bibitem{Tandon20049} Tandon S, Beleggia M, Zhu Y and De Graef M 2004 {\it J. Magn. Mag. Mater.} {\bf 271} 9
\bibitem{PhysRev.71.270.2} Kittel C 1947 {\it Phys. Rev.} {\bf 71} 270
\bibitem{Schlomann} Schlomann E 1964 {\it J. Appl. Phys} {\bf 35} 159
\bibitem{PhysRevLett.88.047204} Jorzick J, Demokritov S O, Hillebrands B, Bailleul M, Fermon C, Guslienko K Y, Slavin A N, Berkov D V and Gorn N L 2002 {\it Phys. Rev. Lett.} {\bf 88} 047204
\bibitem{JPhysChemSolids.11.315} Rado G T and Weertman J R 1959 {\it J. Phys. Chem. Solids} {\bf 11} 315
\bibitem{PhysRevB.72.014463} Guslienko K Yu and Slavin A N 2005 {\it Phys. Rev. B} {\bf 72} 014463
\bibitem{Fischbacher} Fischbacher T, Franchin M, Bordignon G and Fangohr H 2007 {\it IEEE Trans. Mag.} {\bf 43} 2896
\bibitem{PhysRevB.64.144421} Ebels U, Duvail J -L, Wigen P E, Piraux L, Buda L D and Ounadjela K 2001 {\it Phys. Rev. B} {\bf 64} 144421    
\bibitem{PhysRevB.70.054409} Grimsditch M, Giovannini L, Montoncello F, Nizzoli F, Leaf Gary K and Kaper Hans G 2004 {\it Phys. Rev. B} {\bf 70} 054409
\bibitem{Zhang} Zhang Z, Hammel P C and Wigen P E 1996 {\it Appl. Phys. Lett.} {\bf 68} 2005
\bibitem{Klein} Klein O, de Loubens G, Naletov V V, Boust F, Guillet T, Hurdequint H, Leksikov A, Slavin A N, Tiberkevich V S and Vukadinovic N 2008 {\it Phys. Rev. B} {\bf 78} 144410 
\end{thebibliography}
\end{document}